\documentclass[a4paper, 11pt, notitlepage]{scrartcl}
\usepackage[margin=1in]{geometry}                
\usepackage[parfill]{parskip}    

\usepackage[T1]{fontenc}
\usepackage{libertine}

\usepackage{amsmath}
\usepackage{amssymb}
\usepackage{graphicx}
\usepackage{setspace}
\usepackage{xcolor}
\usepackage{authblk}
\usepackage{epigraph}

\newcommand{\be}{\begin{equation}}
\newcommand{\ee}{\end{equation}}
\newcommand{\bes}{\begin{eqnarray}}
\newcommand{\ees}{\end{eqnarray}}

\title{\bf Cryptographic Nature
}

\date{}                                           

\author[1,2]{David Krakauer}

\affil[1]{Santa Fe Institute, Santa Fe, USA}
\affil[2]{Wisconsin Institute for Discovery \\ University of Wisconsin \\ Madison, USA}

\begin{document}
\maketitle

\epigraph{"Nature loves to hide" \\ Heraclitus. }

\begin{abstract}
I consider the many ways in which evolved information-flows are restricted and metabolic resources protected and hidden -- the thesis of living phenomena as evolutionary cryptosystems. I present the information theory of secrecy systems and discuss mechanisms acquired by evolved lineages that encrypt sensitive heritable information with random keys. I propose that complexity science be considered a cryptographic discipline as ''frozen accidents'', or various forms of regularized randomness, historically encrypt adaptive dynamics.
\end{abstract}

\section*{Secret life}

The principle of energy conservation underpins much of physics and chemistry, to include mechanics, thermodynamics and relativity \cite{Anonymous:qYv6cL3l}. By contrast the idea of statistical information provides a common language for genetics, development and behavior, where information encodes ''coordinates'' for free energy landscapes -- memory required by search strategies directed at appropriating metabolic resources \cite{Krakauer:2011wo}. This memory is often encoded in genetic sequences that express enzymes, signaling molecules, and receptors, all of which can orient towards or process metabolic free energy. Whereas our understanding of energy can be reduced ultimately to symmetry principles  \cite{KatherineBrading:2003uo}, information is derived from its antithesis, symmetry breaking \cite{GellMann:1995tg}. Stored information records which of several alternative adaptive configurations have been selected historically.

Adaptive matter requires metabolic energy for biological work -- growth, reproduction and repair \cite{Lotka}. These requirements extend upwards to populations, ecosystems and even cities. The patchy distribution of free energy in space and time, combined with the fact that energy is transformed and thermalized when metabolized, requires efficient and accurate procedures for energy seeking. This scarcity problem is the reason why life can be thought of as a suite of evolved inferential mechanisms dependent on both memory storage -- information -- and ''computation'' for adaptive search. 

Evolution by natural selection -- a population-based search mechanics -- produces outcomes in which information is restricted  and concentrated ensuring that energy is distributed non-uniformly and anti-competitively: sequestered within cells, bodies and communities, where it can be preferentially utilized and monopolized \cite{Krakauer:2009cm} .

Organizations -- cells, organisms and populations -- with accurate information about the whereabouts of metabolic energy sources, endeavor to keep this information to themselves and share informative signals only with those with whom they have found a means to cooperate. These preferred signals include those generating the immunogenic self versus non-self, mating types, restriction systems, species isolating mechanisms, and in some cases, languages \cite{krakinfo}. 

The best way to restrict the flow of information is to protect it or to encrypt it \cite{Shannon:1998vh}. This is the subject of this paper, the many ways in which evolved information-flows are restricted and metabolic resources protected and hidden, the thesis of living phenomena as evolutionary cryptosystems.

\section*{Imitative entropy}

What does it mean for life to be a secret? If it is a secret it seems to be a secret like Edgar Allan Poe's purloined letter, a secret that is hiding in plain sight \cite{Poe:1976ts}. 

One way to think about adaptive secrecy is that it is a necessary step in the evolution of life - protolife and ongoing -- ensuring that replicative and metabolic mechanisms and products are concentrated and protected. Fidelity and privacy are required for continued replicative kinetics. 

The replicator-mutator equation \cite{Nowak:2006wy} describes the frequency of strains $x_i$ in an evolving population of $n$ strains subject to mutation $Q_{ij}$ and selection $k_i$. 

$$ \dot{x}_i = \sum_j Q_{ij}k_jx_j - x_i\bar{k}(\bf x) $$

The operator $Q$ is a bistochastic matrix of transition probabilities. The dynamics are restricted to the simplex $S_n, \sum_i  {x(t)}_i=1$, and the population mean fitness is defined as $ \bar{k}(x(t))$. Under the operation of $Q$, trajectories flow from outside of the positive orthant into $S_n$ with new strains emerging through mutation that are not present in the initial distribution of $x_i(0)$. It is a classical result of evolutionary dynamics that the elements of the $Q$ matrix select between ergodic and non-ergodic equilibrium states; if the off-diagonal elements of $Q$ sum to a value greater then a critical threshold value (''error threshold''), then the distribution of $x_i(t) \rightarrow 1/n  \cite{Eigen:2000ts}.$

The vector $k_i$ encodes the information that a strain $X_i$ possesses about the environment in which it lives. This information is assumed to translate directly into its instantaneous rate of growth $k_i x_i$ and instantaneous relative fitness $x_i(k_i -\bar{k(\bf x)})$. Such selection equations tell us nothing about the adaptive mechanisms through which the information in $\bf k$ is acquired. This requires an adaptive dynamics for $ \bf k$. Under a purely positive imitative dynamics in which $k_i > 0$ is information and $k_i = 0$ is zero information (ignoring the original source of the information), we can include learning, 

$$ \dot{k}_i = x_i \sum_j c_{ij} x_j (\frac{1}{2}\tanh(g(k_j-k_i)+1) - ek_i $$
where $c_{ij}$ is a matrix of imitation rate constants, $g >>1$, and $e$ is an entropic parameter that introduces the loss of information. When information is barely lost $e \gtrapprox 0$ and $Q_{ii} = Q_{jj}$ for $\forall_j \in n$ (all strains are equally mutable) and $c_{ij} = c$,  this learning rule ensures that all strains evolve to a constant quantity of information and the population will converge on $x_i \rightarrow 1/n.$ Thus without some form of secrecy, evolution becomes ''neutral'' as a result of imitative entropy much the same way that evolution becomes a pure drift process above a dissipative error threshold. For evolution to proceed we need to place restrictions - encryption -  on the imitation rate matrix $\bf c$. In order to understand constraints on $\bf c$ we consider in more detail the structure of individual strains in terms of combinatorics and information theory. 

\section{The Communication theory of secrecy systems}

The essential elements of a secrecy system are a message (sometimes call the plaintext), an encrypter that makes use of a key to transform the message into a cryptogram or cipher text. The cryptogram is transmitted -- whereupon it can be intercepted -- to a decrypter that yields the original message using prior knowledge of the key \cite{Shannon:1998vh}. 

For example a ''one-time pad'' is a method of encryption where each bit of a plaintext is encrypted by combining it with the corresponding bit or character from a pad/key using modular addition. The key is random and hence of the same length as the message.  Since every cryptogram is associated with a unique random key, the cryptogram can not be deciphered -- they are information-theoretically secure. 

Shannon proved that information theoretic security is a general property of any effective probabilistic cryptogram.  Define the entropy of a message source as $H(M)$, the entropy of keys $H(K)$ and the entropy of cryptograms as $H(C)$. The desirable properties of an encrypted message can be stated in terms of a triplet of inequalities using mutual information: 

\begin{enumerate}

 \item There is no information in the messages about the cryptograms, $I(M,C) = 0.$

 \item The information in the cryptogram about the key is greater than or equal to zero,  $I(C,K) \ge 0.$

 \item The entropy/uncertainty in the key must be greater than or equal to the entropy/uncertainty in the message, $H(K) \ge H(M)$

\end{enumerate} 

This is the requirement for ''theoretical security'' meaning the cryptogram can never be decoded even with unlimited computing resources. Another way to present this idea is in terms of combinatorics \cite{Massey:1986ty}. 

Consider a set of messages $M$. This set consists of $2^l$ binary sequences of length $n$, where $n \ge l$. The one-time pad has a key with length $n$ where all $2^n$ keys are equally probable. We calculate the cryptogram through modular addition, 
$$ C =  K \oplus M $$
where $\oplus$  is addition modulo 2. This can be shown to ensure that $P(C = c|M=m)  = 2^{-n} $. The probability of discerning a given cryptogram for a given message is strictly chance and the message and the cryptogram are statistically independent. Hence the key must be of the same dimension as the message when a unique key is used to send a single message. 

When a potential attacker has prior information, the demands on the key grow exponentially. For example, if we assume that the attacker has almost perfect knowledge of the crypto system with a memory of $2^l-2$ message-cryptogram pairs, then one requires $2^n$ bits of key information per individual bit in the message. 

The take-home message is that secrecy requires that the information content of a key exceed the information content of a message. In terms of  imitative entropy, this implies that any adaptive system vulnerable to an imitation-exploit requires for ongoing evolution an equal quantity of randomizing information as functional information.

\section{Functional requirements for evolutionary encryption}

The only sure way an evolving lineage can overcome imitation-exploits is to effectively zero its vulnerable imitation coefficients in the matrix $\bf c$. In order to do this, any visible component of an adaptive strategy needs to be mathematically composed with an effectively random key in order to generate an adaptive ''one-way-function''. A one-way-function is a function that is computationally hard to invert given a random input \cite{Katz:2007wa}. For a key-encrypted message such a function, $C = F_K(M)$ has the property that finding $M$ by inverting $F_K$ requires an exhaustive search without the key. Note that once this inversion is found it will be of immediate selective value to an imitator. 

Selective one-way-functions can be generated a number of different ways. Pseudorandom generators provide algorithms for generating sequences that are indistinguishable from random. Pseudorandom permutations build on pseudorandom functions to generate permutations of sequences in memory with uniform probability. A common example of a Pseudorandom permutation would be a block cipher which involves an encryption function, $F_K(M)$ and an inverse decryption  function $F^{-1}_K(C)$. The encryption function breaks information into two blocks of length $k$ and $m$ for key and message, and generates a cryptogram of length $n$,
$$ F_K(M) := F(K,M):\{0,1\}^k \times \{0,1\}^m \rightarrow \{0,1\}^n $$
The decryption achieves the inverse,
$$ F^{-1}_K(C) := F^{-1}(K,C):\{0,1\}^k \times \{0,1\}^m \rightarrow \{0,1\}^n,$$
such that
$$ \forall_K:F_K^{-1}F_K(M)=M.$$
This typically implies that for every unique choice of key one permutation from a set of size  $(2^n)!$ is selected. 

There are a variety of algorithms in computer science that implement block-ciphers, of interest here are those that plausibly map onto candidate adaptive one-way-functions. 

\section{Candidate adaptive one-way-functions}

In this section I review a small number of naturally occurring ''eco-algorithms'' \cite{Valiant:2013vh} that generate a significant amount of randomness in biological sequences that prevents some form of natural ''signal detection''. In a few cases, the function of randomization is not known, in others, there is more or less direct evidence for a form of encryption that achieves protection through reduced competition. It is evident however that randomization in the service of crypsis is ubiquitous in evolved organisms even if the encryption is approximate and far from the limit of ''information theoretic'' security.

\subsection{Shuffled parasites} 

The best studied mechanisms for generating random variation in order to confuse would be surveillants comes from antigenic variation in endoparasites. Parasites that live within long-lived hosts expose their surface proteins to the innate and adaptive immune system. Once recognized,  parasites can be neutralized by antibodies that target exposed markers (antigens) and induce a panoply of clearance mechanisms including agglutination and phagocytosis. It is in the interest of parasites to present ''noise'' to their hosts and thereby evade surveillance. 

The Protozoan, {\em Trypanosoma brucei} causes sleeping sickness and is covered in a protective ''variant surface glycoprotein'' (VSG) which acts somewhat like a cloak of invisibility from the immune systems \cite{Hutchinson:2007bc}. Within a single clone of a single infection it is not unusual for more than $100$ VSG strains to be co-circulating. In order for the parasite to be transmitted to a new host at least one of these strains needs to evade detection. The VSG ''key'' is composed of one protein from a reservoir of between 1000 and 2000 ''mosaic gene'' sequences carried on about 100 mini-chromosomes (estimated to be around 10\% of the total genetic information). Every cell generation a new gene from which the VSG is translated is activated stochastically with a low probability ($ p\approx 0.01$).  Moreover, the mosaic genes -as the name suggests - are generated through stochastic sequence recombination.  Thereby trypanosomes give the appearance of randomness to each immune system that they encounter and reduce the ability of immune memory to mount an effective response to new infections. Comparable randomizing mechanisms are found broadly across the protozoans and the fungi \cite{Verstrepen:2009jh}.

\subsection{Combinatorial ciliates}

The ciliates are an extremely diverse class of unicellualr organism which includes the the freshwater and free-living {\emph Paramecia} and {\emph Tetrahymena}, and the giant  {\emph Stentor} or Trumpet animalcule. Ciliates have two defining characteristics, they are fringed by small hair-like cilia, giving them under the microscope the appearance of independent-minded eyelids fringed by beautiful eyelashes, and they contain two genomes - a small micronucleus (MIC) and a large macronucleus (MAC).

The MIC acts as the reproductive germline and is passed from parent to progeny whereas the macronucleus (MAC) performs the somatic functions of transcription and translation required for growth and asexual cell division. The MAC is generated by the MIC at the start of each new generation through an incredibly complicated sequence of genetic rearrangements, concatenations and amplifications \cite{Prescott:2000kc}. If the MAC is Melville's Moby Dick (around 200,000 words), then the MIC is T.S. Eliot's The Waste Land (around 3,000 words). We further require that Moby Dick is written through an ingenious shuffling and deletion(!) of content from The Waste Land. 

The ability to  construct the MAC requires unshuffling the apparently random MIC into highly ordered genetic sequences using a variety of epigenetic mechanisms including RNA interference (RNAi) and RNA-binding proteins. In some ciliates, around 90\% of the spacer DNA in the MIC is destroyed in preparing the MAC and this seems to play the role of a ''cipher key'' in allowing the adaptive sequence required by the MAC to be decoded each new generation \cite{Bracht:2013ht}.

Unlike the scrambling of the immunogenic surface proteins by protists it is not known why the MIC of the ciliates is so highly randomized. The discoverer of these mechanisms David Prescott has suggested that adaptive variability supporting ''evolveability'' could play a role \cite{Prescott:2000kc}. The hypothesis that I favor is that of encryption that keeps critical information required for replication away from parasites such as viruses that might otherwise appropriate MIC genes for their own translational objectives. For a virus to steal from the MIC it would need to discover and encode the full ciliate decryption function, $F^{-1}_K(C)$ which given the size constrains placed on viruses, is effectively impossible.

\subsection{The genetics of speciation}

Species are defined in many different ways but one feature shared by all definitions is a reduced rate of genetic recombination between species over that within a species. The theory that best accounts for this empirical regularity was provided by Dobzhansky (1936) and then Muller (1939) \cite{Orr:1996bd}. Their idea was that two genetic loci shared by two diploid organisms (\emph{aa, bb}) diverge into two incompatible genomes, (\emph{Aa, bb}) and (\emph{aa, Bb}). The genomes (\emph{AA, bb}) and (\emph{Aa, bb}) remain viable whereas the mixtures, (\emph{Aa, Bb}) do not.

The alleles \emph{A} and \emph{B} have not previously met and interact deleteriously. It is now known that numerous so called ''complementary'' loci play a role in promoting incompatibility between organisms and that many of the alleles that confer incompatibility do little else but promote species isolation. In other words ''speciation genes'', or perhaps more accurately speciation-networks, behave like cryptographic keys ensuring that genetic information conferring locally adaptive information cannot be decoded upon recombination. As Orr has shown these ''complementary'' loci ''cryptographic-loci'') are expected to accumulate at an exponential rate through a mutation-selection process, and so in time, genomes will house a very significant quantity of encryption \cite{Orr:1995ta}.

\section{Cryptographic complexity}

An immediate implication of ''cryptographic nature'' is that a considerable quantity of stored information is required to generate effective keys. The more historically knowledge competitors have stored about the correlation of messages to cryptograms the greater the fraction of key-related resources are required.

By analogy with engineered crypto systems, it is anticipated that key-related random, information will grow and possibly overshadow functional information. This has the implication of creating a rather perverse scaling law. As the quantity of pragmatic adaptive information increases (information of inferential and metabolic value to an organism), the quantity of key-related information increases exponentially. Thus apparently ''junk'' sequences could grow to exceed coding sequences, and this effect will be most pronounced in those lineages with the most functional information. 

\subsection{The''c-value'' paradox}

It has been recognized for several decades that variation in genome size does not correlate in an obvious way with variation in organismal phenotypes. For example, some flowering plants contain up to $10^{11}$ base pairs, whereas some insects have as few as $10^{8}$ base pairs and mammals have not been measured to exceed  $10^{10}$ base pairs. This variation seems to be at odds with the feature sets of these lineages -- why should small, sedentary plant genomes exceed the genome size of large and behaviorally complicated mammals? The''c-value'' (''characteristic value'') paradox describes the surprise biologists feel when confronted with these facts \cite{Eddy:2012js}.

The dominant explanation for the c-value anomaly is that genomes contain large quantities of junk DNA: DNA that has not been hitherto implicated in coding for functional proteins. This has long been thought to reflect the replicative efforts of selfish transposons -- sequences that endlessly ''copy and paste'' themselves throughout a genome with impunity. However, the recent project ENCODE finds that more than $80\%$ of junk is actively transcribed and this at least challenges the most naive selfish replicator model. 

Of significant interest has been the discovery of a veritable zoo of regulatory RNA molecules \cite{Dandekar:1998vu} and conserved non-coding sequences (CNS) \cite{Dermitzakis:2005kda}. Both of these classes of genetic sequence are known to play a crucial role in regulating their less numerous coding-cousins. 

The ''c(ryptographic)-value'' hypothesis suggests that this non-coding sequence could play the role of a key to obscure adaptive information that might otherwise be appropriated by eavesdroppers such as parasites. Comparative studies on bacterial genome variation suggests a possible role for parasitism in inducing an expansion in genome size \cite{Wernegreen:2005bg} and correspondingly a reduction in genome size attendant upon adopting mutualistic relationships \cite{Krakauer:2002wm}.

\section{Complexity science as cryptographic inquiry}

Theoretical science seeks to achieve a compression. Variability in replicated observations is accounted for in terms of minimal models that are: (1) explanatory -- to imply inter-operable with models for nominally distinct  observations often at more microscopic levels, (2) predictive -- capable of producing verifiable outputs for inputs not in the domain of model construction, namely out of sample fidelity, and (3) comprehensible -- revealing of causal relationships between observables. The phrase ''mechanistic model'' stresses properties 1 and 3. Statistical models stress property 2. Phenomenological models stress properties 2 and 3. 
The more fundamental an observation the more successful we have been at achieving all three goals through  minimal models. The theory of electromagnetism would be a prototypical example.

In the domain of complex phenomena we struggle to obtain all of these properties in a minimal model.  All physical systems can be decomposed into regular and noisy factors, but complex systems are typified by an unusually high contribution from randomness. And this randomness can be of a special kind that introduces new forms of regularity.  Forms of probabilistically fixed randomness ($p=1$) that produce new regularities allow for a direct connection to encryption.

\subsection{Frozen accidents}

To make the role of ''regularized randomness'' or  ''frozen accidents'' \cite{GellMann:1995tg} clear, consider the world of one-dimensional cellular automata (CA). The CA is characterized by a 5-tuple, 

\begin{equation}
C = (S, s_0, N, d, f)
\end{equation}

The value $S$ is the set of states and $s_0$ the initial state of the CA. The neighborhood is $N$, the dimension $d$, and the local update rule, $f:S^i \rightarrow S $. At each point in time the global system is updated, $F: C_t \rightarrow C_{t+1}$. We shall consider $d=1$ and a binary state space, $S \in \{0,1\}$. The neighborhood is defined by a symmetric radius, $r=1$ -- nearest neighbors, in which case $N = \{-r,0,r\}$ which implies a neighborhood size of $3$.

For this kind of CA there are $256$ update rules given by the state of three binary random variables, $p,q,r$. For example rule 240 depends only $p$. Hence this rule is indifferent to the values of $q$ and $r$. An initial state $s_0 = (p=1, q=0, r=0)$ generates a global pattern that resembles a diagonal line oriented at 45\% to the right of vertical. As does any initial state $s_0 = (1**)$. At some time $\tau$ the system will possess a state $S_{\tau}$. 

Now assume a stochastic dynamic that operates on the space of update rules,  $D(i,j):f^{(i)} \rightarrow f^{(j)}$. Where $i$ indexes a rule $f^{(i)}$ in the set of $256$ possible rules. We might choose that rules can ''mutate'' into another as a function of their absolute distance in rule-space with some probability $p<<1$. Hence 

\begin{equation}
D(i,j) = p^{|i-j|}
\end{equation}

Each new mutation is itself unpredictable but the consequence of each mutation is to change the deterministic laws of motion. The observed system $C^{(i)}$, where $i$ indexes the current update rule, will randomly walk in a discrete lattice of highly regular dynamics. For example, if we start with rule $240$ we might observe one realization of this random walk to be: $241$, $243$, and return to $242$. This corresponds to a motion in the space of boolean functions: 

\begin{equation}
p \rightarrow p \lor (\lnot(q \lor r))  \rightarrow p \lor (\lnot q) \rightarrow p \lor (r \land((\lnot q))) 
\end{equation}

This illustrate the role that an evolutionary process can have on the prediction of dynamics from a given initial condition, $s_0$.   Mutations break the symmetry of sites $q$ and $r$. Without knowledge of the precise transitions, this message has been effectively encrypted through the function 
$D(i,j)$. In order to predict the evolution of the system from the initial condition, we would need to know both the time at which mutation took place and the selected rule.

\subsection{Historical encryption}

 Complexity science can be likened to efforts at decoding a message that has been combined with a variety of historical sources of noise. Note that the noise needs to remain low in order that the original signal persists. The degree of minimalism that any denoised model can achieve will then be captured through concepts such as ''effective complexity'' \cite{Lloyd:1996wg}. 

The effective complexity of a complexity theory, to the extent that it seeks to fit observed regularities, is expected to be rather large. Even a parsimonious, predictive model will need to take into account a known historical key that accounts for each branching event upon which a symmetry has been broken.  Unlike human crypto systems systems, evolved keys have not in any deliberative way been transmitted and the encryption is not ''designed'' to conceal the nature of reality, this arises as an accident of history promoting a switch between different dynamical scenarios -- hence ''frozen accidents''.

One way in which we might decode dynamics in the scientific search space is to consider complex systems at the level of universality or ''equivalence classes'' under coarse-grained measurements.  

\subsection*{Compressed historical cryptographic keys}

One means of reducing the quantity of historical key is to recognize that conservation laws impose significant constraints on the space of realizable forms and functions. All adaptive solutions need to be consistent with what we know of physics and chemistry. 

For the example from CAs we described the random realization:

\begin{equation}
p \rightarrow p \lor (\lnot(q \lor r))  \rightarrow p \lor (\lnot q) \rightarrow p \lor (r \land((\lnot q))). 
\end{equation}

This corresponds to a random walk in the space of coarse-grained Wolfram universality classes (conventionally numbered 1-4) \cite{Wolfram:1984fn}: 

\begin{equation}
2 \rightarrow 2 \rightarrow 2 \rightarrow 2.
\end{equation}

In other words, this random sequence of deterministic cases is invariant with respect to its statistical evolution which preserves patterns in which we find a set of separated simple stable or periodic structures (Class 2).  This need not be true, for different rules, we could observe at the ''macroscopic'' level transitions between universality classes. However, through a higher level of description we can often significantly increase our predictive capability. The same logic is pursued in natural science. 

The theory of allometric scaling and regular features of evolved networks provide a means of compressing contingent information by exploiting constraints on metabolic energy optimization. It is well known that all multicellular organisms conform to scaling laws with quarter power exponents \cite{West:1999tm}. 

At the genetic level it is  observed that mutation rate scales with mass according to $ p = km^{-1/4}$. If we assume that organisms seek to maximize the rate of mutation subject to the upper limit of the error threshold, then we can derive a maximum genome size, $L = k’m^{1/4}$ and thereby connect adaptive information directly to metabolic free energy, $L=k'B^{1/3}$ \cite{Krakauer:2011wo}. Hence without any knowledge of history we can place an upper bound on the maximum dimension of the genome that can be propagated forward in time under the most optimistic distribution of free energy. This does not tell us which sequence an organism is using but it does define a metabolic equivalence class for all organisms with a given quantity of genetic information.  This kind of prior information makes decrypting evolutionary history somewhat easier but significant challenge remains.

The same kind of argument, albeit somewhat weaker, comes from network science. There is evidence for empirical laws that describe the distributions of network connectivity \cite{Newman:2006daa}. The degree distributions of many evolved networks to include genetic regulatory networks, neural networks and ecological networks fall into a rather small number of functional forms to include power laws. Higher order properties of these networks include the logarithmic scaling of geodesics with network size - so called ''small world networks'' \cite{Watts:1998vc}. As in the case of allometry, these regularities provide us with prior information that we can use to restrict the search space of historical keys - in this case network growth rules or developmental processes that are capable of producing the observed statistical features of the networks.

\section*{Conclusion: from invariance to complexity}

Eugene Wigner in his influential paper, ''Events, laws of nature, invariance principles'' \cite{Wigner:1964et} maintains that it is the role of physics to account for those regularities that are called ''laws of nature''. The elements of the behavior which are not specified by the laws of nature are called initial conditions''. It is by the definition of initial conditions that they are arbitrary with respect to physical laws, but complex phenomena accumulate an alarming quantity of initial conditions and they are far from functionally arbitrary. This places us in a pickle of a problem. 

One way to reduce the magnitude of this challenge is to restrict the volume of initial conditions by using our knowledge of physics and biophysics. Allometry, network regularities and error thresholds are all principled approaches to placing bounds on free parameters, using geometric invariance principles \cite{Krakrock}, to limit the size of a theory. 

Another approach suggested in this paper is to begin to formalize the complexity science of adaptive systems in terms of mathematically suggestive concepts from the field of cryptography. This resembles ''metamathematics'' -- the mathematical study of mathematical objects -- in that we are endeavoring to formalize generic informational properties that result from any evolutionary, communicative, process. In order to explain specific cases we have no choice but to significantly increase the number of measured parameters and our theories and models are anticipated to grow accordingly.

\bibliographystyle{siam}

\end{document}